\documentclass[11pt]{article}
\usepackage{amsmath}
\usepackage{amssymb}
\usepackage{hhline}
\usepackage[scale={.75,.75}]{geometry}
\usepackage{url}
\usepackage{caption}
\usepackage[numbers, sort&compress]{natbib}
\usepackage{epsfig}
\usepackage{multirow}
\usepackage{color}
\usepackage{verbatim}
\usepackage{nicefrac}
\usepackage{upgreek}
\usepackage{hyperref}
\usepackage{bbm}
\usepackage{setspace}

\newcommand{\slashed}[1]{\displaystyle{\not}{#1}}

\begin{document}
\date{}
\title{ 
{\normalsize TTK-12-04\hfill\mbox{}\\}
\vspace{0.5cm}
{\bf \boldmath 
Matter and Antimatter in the Universe}\footnote{Prepared as invited contribution to \cite{FocusIssue}}}
\author{Laurent Canetti$^{a}$, 
Marco Drewes$^{b,c}$, 
Mikhail Shaposhnikov$^{a}$\\ \\
{\normalsize \it$^a$ ITP, EPFL,
CH-1015 Lausanne, Switzerland}\\
{\normalsize \it$^b$ Institut f\"ur Theoretische Teilchenphysik und Kosmologie,} \\{\normalsize \it RWTH Aachen, D-52056 Aachen, Germany}\\
{\normalsize \it$^c$ Physik Department T31, Technische Universit\"at M\"unchen,} \\{\normalsize \it James Franck Stra\ss e 1, D-85748 Garching, Germany}
}
\maketitle
\begin{abstract}
We review observational evidence for a matter-antimatter asymmetry in
the early universe,  which leads to the remnant matter density we
observe today.  We also discuss  bounds on the presence
of antimatter in the present day universe, including the possibility
of a large lepton asymmetry in the cosmic neutrino background. We
briefly review the theoretical framework within which baryogenesis,
the dynamical generation of a matter-antimatter asymmetry, can occur. 
As an example, we discuss a testable minimal particle physics model
that simultaneously explains the baryon asymmetry of the universe,
neutrino oscillations and dark matter.
\end{abstract}

\section{Introduction}
The existence of antimatter is a direct consequence of combining two
of the most fundamental known concepts in physics, the theory of
relativity and quantum mechanics.  Its theoretical prediction, based
on these abstract principles \cite{Dirac:1928hu}, and experimental
discovery \cite{Anderson:1932zz} mark one of the great successes of
theoretical physics.  At the time of its discovery, antimatter was
thought to be an exact mirror of matter; all phenomena that had been
observed in nature were invariant under conjugation of parity (P) and
charge (C) as well as time reversal (T), and not much was known about
the early history of the universe.  Henceforth, the enormous
matter-antimatter asymmetry of the nearby universe (complete absence
of antimatter except in cosmic rays) posed a mystery that could only
be explained by assuming that the universe was set up like this.

With rise of the big bang theory after the theoretical prediction
\cite{Friedman,Lemaitre} and observational discovery  of the cosmic
expansion \cite{Hubble} and microwave background (CMB)
\cite{Penzias:1965wn}, it came clear that the universe was hot during
the early stages of its history \cite{Gamow:1946eb}, and antimatter was present when pair-creation and annihilation reactions were in thermal equilibrium. When particle energies in the cooling plasma became too small for pair creation to occur, almost all particles and antiparticles annihilated each other, with a small amount of matter (by definition) surviving. 

The baryon asymmetry of the universe (BAU) can
be defined as the difference between the number of baryons $N_B$ and antibaryons $N_{\bar{B}}$ divided by their sum (or the entropy $s$) just before antiprotons disappeared from the primordial plasma.  Since the end products of annihilation processes are mostly photons and there are no antibaryons in the universe today \footnote{Here we assume that the BAU is the same everywhere in space within the observable universe, we discuss this point in section
\ref{sec:AMinUniverse}.} , the BAU can be estimated by the baryon to photon ratio $\eta$,
\begin{equation}\label{etadef}
\eta=\frac{N_B}{N_\gamma}\Big|_{T=3K}=
\frac{N_B-N_{\bar{B}}}{N_\gamma}\Big|_{T=3K}
\sim\frac{N_B-N_{\bar{B}}}{N_B+N_{\bar{B}}}\Big|_{T\gtrsim1{\rm GeV}}
.\end{equation}
In the last term we have expressed the temperature in GeV, with $1 {\rm GeV} \simeq 1.16\cdot 10^{13}{\rm K}$.
$\eta$ is related to the remnant density of baryons $\Omega_B$, in
units of the critical density, by
$\Omega_B\simeq\eta/(2.739\cdot10^{-8}h^2)$, where $h$ parameterises
the the Hubble rate $H_0=100 h$ (km/s)/Mpc. It can be determined
independently in two different ways, from the abundances of light
elements in the intergalactic medium (IGM), see section \ref{sec:BBN},
and from the power spectrum of temperature fluctuations in the CMB,
see section \ref{sec:CMBLSS}. Both consistently give values
$\eta\sim 10^{-10}$; the precise numbers are given in section
\ref{sec:EtaMeasure}. Thus, today's huge matter-antimatter asymmetry 
was actually a tiny number in the past.  The discovery of
violations of P \cite{Wu:1957my} and CP \cite{Christenson:1964fg}
invariance (and thus also C invariance) in nature provided hints that this
asymmetry may have been created dynamically by {\it baryogenesis} from
a matter-antimatter symmetric initial state.

There are three necessary conditions for successful baryogenesis,
which were first formulated by Sakharov
\cite{Sakharov:1967dj}\footnote{See also \cite{Kuzmin:1970nx} for a
related early discussion.}: I) baryon number violation, II) C and CP
violation and III) a deviation from thermal equilibrium. Intuitively,
these conditions can easily be understood. Without baryon number
violation, it is not possible for any system to evolve from a state
with baryon number $B=0$ to a state with $B\neq 0$. If C (or CP)
symmetry were to hold, for each process that generates a
matter-antimatter asymmetry, there would be a C (or CP) conjugate
process that generates an asymmetry with the opposite sign and occurs
with the same probability. Finally, thermal equilibrium is a time
translation invariant state in which the expectation values of all
observables are constant, therefore it requires a deviation from
equilibrium to evolve from $B=0$ to $B\neq 0$. Formally, Sakharov's
conditions can be proven by means of quantum mechanics and statistical
physics. We describe the universe as a thermodynamic ensemble,
characterised by a density matrix $\varrho$.  In the Schr\"odinger
picture, $\varrho$ evolves in time according to the von Neumann (or
quantum Liouville) equation
\begin{equation}
i\frac{\partial \varrho(t)}{\partial t}=[{\rm H},\varrho(t)]
,\end{equation}
where ${\rm H}$ is the Hamiltonian. The baryon number is given by
$B(t)={\rm tr}({\rm B}\varrho(t))$, where ${\rm B}$ is the baryon
number operator. If $[{\rm B},{\rm H}]=0$ and $B=0$ at initial time,
then $B=0$ at all times, which proves I). To prove II), we consider an
arbitrary discrete transformation ${\rm X}$ that commutes with ${\rm
H}$ and anticommutes with ${\rm B}$.  If $\varrho(t)$ at some time
$t_0$ is symmetric under ${\rm X}$, i.e. $[{\rm X},\varrho(t_0)]=0$, then
this holds for all times. Thus, in order to create a CP-asymmetric
state from a symmetric initial state, ${\rm H}$ must not commute with
${\rm CP}$. The proof of III) is trivial since in thermal equilibrium,
$\varrho^{eq}$ is time translation invariant by definition, thus $B$
is constant.

The paradigm of {\it cosmic inflation} \cite{Starobinsky:1980te}
elevated the idea of an initial state with $B=0$ from an
assumption, based on aesthetic reasoning, to a generic prediction.  If
the universe underwent a period of accelerated expansion during its
very early history that lasted for long enough to explain its spacial
flatness and the isotropy of the CMB temperature, any pre-existing
baryon asymmetry was diluted and negligible at the end of inflation
\footnote{If $B$ and $L$ are violated individually, as e.g. in the
model presented in section \ref{nuMSM} or thermal leptogenesis, a
state with $B=0$ is also reached unavoidably, even for an initial
$B\neq 0$, when the universe reaches chemical equilibrium.}. 
Therefore, baryogenesis needs to occur either during reheating or in
the radiation dominated epoch.

The Standard Model of particle physics (SM) and cosmology in principle
fulfills all three Sakharov conditions \cite{Kuzmin:1985mm}. Baryon
number is violated by sphaleron processes
\cite{'tHooft:1976up,Kuzmin:1985mm}, P and CP are violated by the weak
interaction and the quark Yukawa couplings \cite{Kobayashi:1973fv} and
the nonequilibrium condition is fulfilled due to the expansion of the
universe. As it turns out, the values of the CP-violating Kobayashi-Maskawa phase and mass of the Higgs particle suggested by experiments make it extremely
unlikely that successful baryogenesis is possible within the SM. The
CP violation and deviation from equilibrium during electroweak
symmetry breaking are both too small. These aspects are discussed in
more detail in section \ref{sec:BAUinSM}. However, models of particle
physics beyond the SM generally contain many new sources of CP and
possibly $B$-violation, and a large number of baryogenesis scenarios
is discussed in the literature.\\

{\bf The remainder of this article is organised as follows.} The
following two sections are devoted to observational evidence for a
matter-antimatter asymmetry in the universe.  In section
\ref{sec:AMinUniverse} we review bounds on the existence of primordial
antimatter at present time. We also discuss the possibility that there
is no overall asymmetry in the universe, which is composed of regions
dominated by matter or antimatter. We conclude that it is almost
certain that all structures in the observable universe are composed of
matter only\footnote{Here and in the following ``matter'' and
``antimatter'' refer to baryons. We do not discuss the unknown
composition and origin of dark matter (except in section \ref{nuMSM}),
which in many popular models is not related to $\eta$.}.  In section
\ref{sec:EtaMeasure} we adopt that viewpoint and review current
measurements of the asymmetry parameter defined in (\ref{etadef}). In
section \ref{sec:EtaMeasure} we briefly overview theoretical
approaches to explain the BAU, focusing on testable models. Finally,
in section \ref{nuMSM}, we discuss a minimal model, in which all
new particles may be found using present day observational and
experimental techniques.

\section{Antimatter in the present Universe}
\label{sec:AMinUniverse}
The only place in the present day universe where we can directly look
for antimatter is the solar system, where we have visited and
approached various celestial bodies with spacecraft.  It does not
contain any significant amount of antimatter\footnote{Antimatter in
the solar system can also be excluded because it would lead to a
strong signal when annihilating with solar winds
\cite{Steigman:1976ev}.}. We receive direct probes from other parts of
our galaxy in the form of cosmic rays, which have recently been
intensely studied by the PAMELA and FERMI space observatories. These
contain a fraction of positrons and anti-protons, the only primary
source of antimatter found outside the laboratory to date. However,
since the pair creation threshold for these particles is relatively
low, they can be generated by various astrophysical processes and are
expected to be found even in a universe that is entirely made of
matter otherwise.  If heavier antinucleids were found, this would
indicate that there exists traces of antimatter within our own galaxy.
So far, none have been discovered. The lack of findings by the first
mission of the Alpha Magnetic Spectrometer (AMS) allows to conclude
their absence at a level of $10^{-6}$ \cite{Alcaraz:2000ss}. These
bounds are expected to tighten after data from the AMS 02 experiment,
currently mounted on the International Space Station, is released.  Furthermore, in
\cite{Steigman:2008ap} it was argued that the fraction $f$ of
antimatter in the ISM cannot be larger than $f<10^{-15}$ because the
lifetime of antinuclei in the ISM due to annihilations is only 300
years \cite{Steigman:1976ev}.

Upper bounds on the presence of antimatter in other parts of the
universe can be imposed by indirect detection methods.  One can
distinguish two basic scenarios: Either matter and antimatter are
mixed homogeneously, i.e. the interstellar medium (ISM) or IGM are
matter dominated everywhere in space, but contain a certain fraction
of antimatter\footnote{Such a mixing is not possible within individual
stellar systems because the antimatter would have annihilated during
the gravitational collapse that lead to their formation
\cite{Steigman:1976ev}.},  or patches of matter and antimatter
coexist. In both cases one would expect to observe X-rays and
$\gamma$-rays from annihilation processes.

\subsection{A Patchwork Universe}
\label{patchwork}
If the universe is a patchwork of regions that are strongly dominated
either by matter or antimatter,  the question arises what is the
typical size of such regions. The possibility of individual antimatter
stellar systems has been discussed in
\cite{Steigman:1976ev,Steigman:2008ap}, see also
\cite{Stecker:1984wx}.  The absence of annihilation signals from such
stars passing through the ISM allows to conclude that their fraction
in the galaxy is $<10^{-4}$. Since the presence of such systems is
hard to accommodate within a realistic model of galaxy formation, it
is tempting to conclude that it is zero. Similar arguments can be
brought forward against the possibility of clouds of anti-gas or other
isolated objects in the milky way. This viewpoint has been questioned
in \cite{Bambi:2007cc}, see also references therein. However, no
definite conclusions that hint towards the opposite could be drawn. In
\cite{Steigman:2008ap} it was furthermore pointed out that the authors
of \cite{Bambi:2007cc} may have underestimated the annihilation cross
sections at low energies.

This still leaves open the possibility that the universe is a
patchwork of huge distinct regions of matter and antimatter, in the
most extreme case with vanishing baryon number $B=0$ when averaged
over large volumes. If this were the case, these regions would have to
be at least of comparable size as the observable universe
\cite{Cohen:1997ac,Cohen:1997mt}. This conclusion can be drawn from
the measured cosmic diffuse $\gamma$-ray (CDG) background. After
nonlinear structure formation, the matter and antimatter domains may
have been separated by sufficiently large voids in the IGM to suppress
annihilation at the domain walls and avoid a detectable $\gamma$-ray
flux. However, the homogeneity of the CMB does not allow for such
spacial separation before recombination. Hence, matter and antimatter
domains must have been in touch at least between the time of
recombination and the beginning of nonlinear structure formation. The
$\gamma$-rays produced by annihilation during this period would,
though redshifted, still be present today and contribute to the CDG. 
The measured intensity of the CDG allows to conclude that the domains
must at least have a size comparable to the observable universe
\cite{Cohen:1997ac}.

\subsection{A mixed Universe}
The considerations at the beginning of this section strongly constrain
diffuse antimatter within our galaxy. The fraction $f$ of antimatter
on larger scales is constrained by the measured $\gamma$-ray flux from
the IGM. The IGM emits X-rays due to thermal bremsstrahlung in 2-body
collisions. The expected flux of $\gamma$-rays $F_\gamma$ is
proportional to the flux of X-rays $F_X$. This allows to constrain $f$
as \cite{Steigman:2008ap}
\begin{equation}
f \leq 3\cdot 10^{-11}\frac{T}{\rm keV}\frac{F_\gamma}{F_X}, 
\end{equation}
where $T$ is the gas temperature and the inequality is due to the fact
that not all $\gamma$-rays originate from annihilations. In
\cite{Steigman:2008ap}, the upper bounds on the $\gamma$-flux imposed
by the EGRET space telescope \cite{Reimer:2003er} were used to
constrain $f$ for a sample of 55 galaxy clusters from the limited flux
survey published in \cite{Edge:1990}. The obtained values scatter
between $f<5\cdot 10^{-9}$ and $f<10^{-6}$, indicating that these
clusters consist either entirely of matter or antimatter in good
approximation. Furthermore, if there are any antimatter dominated
regions, they must be separated from the matter domains at least by
distances comparable to the size $\sim$Mpc of galaxy clusters. 
Observations of colliding galaxy clusters allow to extend the analysis
to even larger scales. For the bullet cluster
\cite{Markevitch:2001ri}, an upper bound of $f<3\cdot 10^{-6}$ was
obtained in \cite{Steigman:2008ap}. If representative, this allows to
extend the constraints on $f$ to scales of tens of Mpc. In combination
with the considerations in section \ref{patchwork} this indicates
that  the present day observable universe most likely does not contain
significant amounts of antimatter.

\section{The Baryon Asymmetry of the Universe}
\label{sec:EtaMeasure}
As motivated by the discussion in section \ref{sec:AMinUniverse}, we
in the following adopt the viewpoint that the universe is
baryon-asymmetric and the asymmetry is the same everywhere within the
observable Hubble-volume. Within the concordance model of cosmology
($\Lambda$CDM) \cite{Lahav:2010mi} it can be estimated by the baryon
to photon ratio (\ref{etadef}). There are two independent ways to
determine this parameter, from the relative abundances of light
elements in the IGM on one hand and from the spectrum of temperature
fluctuations in the CMB on the other.  Since they measure $\eta$ at
very different stages  during the history of the universe, they also
provide a check for the $\Lambda$CDM model itself. 

\subsection{Big Bang Nucleosynthesis}
\label{sec:BBN} 
Throughout the evolution of the universe, there was a brief period
during which the temperature was low enough for nuclei with mass
number $A>1$ to exist and still high enough for thermonuclear
reactions to occur.  This {\it big bang nucleosynthesis} (BBN) is
thought to be the main source of deuterium (D), helium ($^3$He,
$^4$He) and lithium (mainly $^7$Li) in the universe
\cite{Gamow:1946eb}, see e.g. \cite{Fields:2006ga} for a review. These
elements, in particular H and $^4$He, make up the vast majority of all
nuclei in the universe\footnote{Heavier nuclei are not produced during
BBN due to the absence of stable nuclei of mass number $5$ and $8$ and
the bigger Coulomb repulsion between for charge numbers $Z>1$. They
can be made in stars (up to iron) or supernovae.}.

The processes relevant for BBN start when the temperature of the
primordial plasma is around $T\sim 2$ MeV with the neutrino freezeout,
i.e. the reactions that keep neutrinos in equilibrium with the plasma
become slower than the expansion of the universe. This energy range is
easily accessible in the laboratory, and the underlying particle
physics is well-understood. Thus, in the standard scenario the sole
unknown parameter that enters BBN is the baryon to photon ration
$\eta$, which was fixed by unknown physics (baryogenesis) at higher
energies.  
For given $\eta$, the time evolution of the different isotopes' abundances can be determined by solving a network of Boltzmann equations.
Thus, within $\Lambda$CDM and the
SM, $\eta$ can be uniquely determined by measuring the primordial
abundances of light elements. The result found in
\cite{Steigman:2010zz}, in units of $10^{-10}$, is
\begin{equation}
\eta_{SBBN}=5.80\pm 0.27,
\end{equation}
more constraining than the 95\% CL value $4.7<\eta_{SBBN}<6.5$ quoted
in \cite{Fields:2006ga}.

The good agreement with observational data allows to impose tight
constraints on theories of particle physics beyond the SM which
predict charged or unstable thermal relics from earlier epoch to be
present around the time of BBN. Even without additional particles that
participate in BBN reactions, non-standard physics may leave a trace
in the abundances of light elements by modifying the expansion rate of
the universe. The expansion rate is given by
\begin{equation}
H^2=\frac{8\pi}{3}G\rho
,\end{equation}
where $G$ is Newton's constant, $H$ the Hubble parameter and $\rho$
the energy density of the universe. In the radiation dominated era at
the time of BBN $\rho\simeq\rho_\gamma+\rho_e+N_\nu\rho_\nu$, where
the first two terms are the energy densities for photons and
electrons/positrons and $\rho_\nu$ is the contribution from a flavour
of neutrinos. $N_\nu$ is the {\it effective number of neutrino
species}. In the standard scenario $N_\nu=3$ during BBN\footnote{At
the time of CMB decoupling $N_\nu=3.046$ in the SM, where the
deviation from $3$ parameterises a deviation from the equilibrium
distribution of neutrinos caused by $e^\pm$ annihilation 
\cite{Mangano:2005cc}.}. Any deviation $\Delta N_\nu$ from that can be
used to parameterise a non-standard expansion rate. Despite the name,
a $\Delta N_\nu\neq 0$ need not be caused by an additional neutrino
species. It may e.g. be due to any non-standard energy budget,
gravitational waves, varying coupling constants or extra dimensions. A
best fit to the observed element abundances with $N_\nu$ left as a
free parameter, reported in \cite{Steigman:2010zz}, yields
\begin{equation}
\eta_{BBN}=6.07\pm 0.33,\ \Delta N_\nu=0.62\pm0.46
,\end{equation}
with $\Delta N_\nu$ consistent with zero at $\sim1.3 \sigma$. 
\footnote{Recently evidence from different sources that hints towards $\Delta N_\nu>0$ after BBN has mounted \cite{Komatsu:2010fb,Hamann:2010bk,Dunkley:2010ge},
 but
the statistical significance does not allow a definite conclusion at
this stage.} However, the precise values of these parameters are
affected by the selection of datasets and estimates of systematic
errors, cf. discussion in \cite{Fields:2006ga}.   

The main uncertainty results from the difficulty to measure the
primordial abundances of light elements.  They differ from present day
values within galaxies, which have been modified by thermonuclear
reactions in stars throughout the past 13 Gyrs. $N_\nu$ is mainly
sensitive to the $^4$He abundance (because the expansion history
determines the point of neutron freezeout, which affects the neutron
fraction $n/p$ in the plasma), $\eta$ is sensitive to D. In contrast
to He, there are no known astrophysical sources of D
\cite{Epstein:1976hq}, thus the observed abundance provides a reliable
lower bound on the primordial value. In fact, the BBN bounds on $\eta$
are almost entirely derived from the D abundance.  This has earned D
the label "baryometer" of the universe, and the strong dependence on D
is the reason why $\eta$ is relatively insensitive to the infamous
$^7$Li problem \cite{Fields:2012jf}.  It is believed that the most
precise measurement to date can be obtained from high redshift low
metallicity quasar absorption systems (QSO), though the systematic
errors are not fully understood, cf. discussion and references in
\cite{Fields:2006ga}.

\subsection{CMB and LSS}
\label{sec:CMBLSS}
The baryon content of the universe can also be determined from the
power spectrum of temperature fluctuations in the CMB.  The
temperature fluctuations were generated by acoustic oscillations of
the baryon-photon plasma in the gravitational potential caused by
small inhomogeneities in the DM distribution. The oscillations are
sensitive to $\eta$ because the baryon fraction determines the
equation of state of the plasma, which mainly manifests in the
relative height of odd and even peaks in the power spectrum.  Since
the decoupling of photons happens at a vastly different epoch
($T\sim0.3$ eV) and the physics that produces the acoustic peaks in
the power spectrum is very different from BBN, this is a truly
independent measurement. The WMAP7 data \cite{Komatsu:2010fb} alone
gives, in units of $10^{-10}$,
\begin{equation}
\eta_{CMB}=6.160^{+0.153}_{-0.156}
.\end{equation}
When the WMAP7 data is combined with large scale structure (LSS) data
\cite{Percival:2009xn} and the Hubble rate $H_0=74.2\pm3.6$km
s$^{-1}$Mpc$^{-1}$ found in \cite{Riess:2009pu}, this value slightly
changes to $\eta_{CMB/LSS}=6.176\pm0.148$ \cite{Komatsu:2010fb}. A
similar analysis, combining different CMB and LSS data sets, was
performed in \cite{Simha:2008zj} and gave results scattered between
$6.1$ and $6.16$ for $\eta$ in units of $10^{-10}$. In the context of
baryogenesis, $\mathcal{O}[1]$ corrections to the above numbers are of
little relevance, given our lack of knowledge of physics beyond the
SM. The impressive agreement between BBN and CMB results for $\eta$,
however, is am important hint that we can reliably determine the
magnitude of the BAU observationally. 

\subsection{A large Lepton Asymmetry?} 
The good agreement between CMB and BBN results strongly constrains the
BAU to be as small as $\eta\sim 10^{-10}$.  Compared to that, bounds on a
lepton asymmetry of the universe (LAU), hidden in the cosmic neutrino
background (CNB), are much weaker. The only source of B-violation in
the SM are sphaleron processes \cite{'tHooft:1976up,Kuzmin:1985mm},
which are highly inefficient below $T_{EW}\sim 140$ GeV, as it would
be suggested by a $126$ GeV Higgs mass \cite{ATLAS:2012ae}.
However, if neutrinos are Majorana particles, the Majorana mass term
can lead to lepton number violating processes at much lower energies.
Furthermore, neutrino mixing may add another source of CP-violation to
the SM. Thus, it is possible to imagine nonequilibrium processes that
generate a LAU below the electroweak scale that is orders of magnitude
larger than the BAU.  Though of little effect today, a large lepton
asymmetry may have far-reaching consequences in the past. It can
trigger a resonant production of dark matter (DM), see section
\ref{nuMSM}, or affect the nature of the QCD transition
\cite{Schwarz:2009ii}.

The main constraints on the LAU come from BBN. It affects BBN in two
ways. On one hand, non-zero chemical potentials modify the momentum
distribution of neutrinos. This changes the temperature dependence of
the energy density and thereby the expansion history. More
importantly, electron neutrinos $\nu_e$ participate in the conversion
processes that keep neutrons and protons in equilibrium. The change in
the freezeout value of $n/p$ caused by a $\nu_e$ asymmetry leaves an
imprint in the $^4$He abundance.

The LAU has been studied by different authors
\cite{Kang:1991xa,Mangano:2011ip,Castorina:2012md}.
It can be defined in analogy to (\ref{etadef}),
\begin{equation}
\eta_\alpha=\frac{N_{\nu_\alpha}-N_{\bar{\nu}_\alpha}}{N_\gamma}\Big|_{T=3K}.
\end{equation}
In absence of neutrino masses, there would be only weak constraints on
$|\eta_\mu|,|\eta_\tau|\lesssim2.6$ (in units of one!) 
\cite{Mangano:2011ip} because at the time of BBN, electrons are the
only charged leptons in the primordial plasma. $\eta_e$ would be
constrained to $-0.012\lesssim\eta_e\lesssim0.005$  due to its effect
on $n/p$. However, neutrino oscillations tend to make the lepton
asymmetries in individual flavours equal\footnote{This process is
sometimes referred to as ``flavour equilibration'', though the
neutrino distribution functions may deviate from Fermi-Dirac because
the process occurs close to the neutrino freezeout and neutrinos may
not reach thermal equilibrium. This means that a lepton asymmetry
cannot be translated into a chemical potential in the strict sense.}
well before BBN \cite{Dolgov:2002ab,Wong:2002fa,Abazajian:2002qx} for
values of the neutrino mixing angle $\uptheta_{13}$ suggested by
experiment \cite{Pastor:2008ti}.  They tie the bounds for all flavours
together and introduce a dependency on the choice of mass hierarchy
and $\uptheta_{13}$.  For $\sin^2\uptheta_{13}=0.04$, at the upper end
of the region suggested by \cite{Fogli:2011qn} and about twice the
value found more recently in \cite{An:2012eh,Ahn:2012nd,Fogli:2012ua}, the bounds
found in \cite{Mangano:2011ip} are
$-0.17\lesssim\eta_{e,\mu,\tau}\lesssim0.1$ for normal and
$-0.1\lesssim\eta_{e,\mu,\tau}\lesssim0.05$ for inverted hierarchy.
For inverted hierarchy, the dependence on $\uptheta_{13}$ is weak,
while for normal hierarchy values as large as
$|\eta_{e,\mu,\tau}|\simeq0.6$ are allowed for small $\uptheta_{13}$.
In \cite{Castorina:2012md}, stronger bounds $-0.071<\eta_{e,\mu,\tau}<0.054$
were found for $\sin^2\uptheta_{13}=0.04$, assuming normal hierarchy,
and it was pointed out that future CMB observations may be able to
compete with BBN in constraining $\eta_\alpha$.


\section{Testable Theories of Baryogenesis}\label{sec:Theory}

\subsection{Baryogenesis in the SM}
\label{sec:BAUinSM}
In principle, the Sakharov conditions I) - III) are all fulfilled in
the SM. This is rather obvious for the conditions conditions II) and
III).  The weak interaction violates P-invariance maximally, while
CP-invariance is violated by the complex phase $\delta_{KM}$ in the 
Cabibbo-Kobayashi-Maskawa matrix. The expansion of the universe brings
the primordial plasma out of thermal equilibrium. The violation of
baryon number, condition I), occurs more subtly\footnote{See
\cite{Shaposhnikov:1991pd} for a pedagogical discussion.}.

\begin{figure}
  \centering
    \includegraphics[width=8cm]{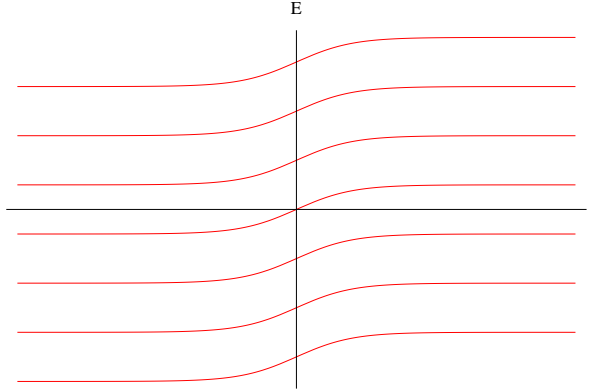}
    \caption{An illustration of fermionic level crossing:  The
    changing bosonic background fields can modify the fermionic energy
    levels, leading to fermion number violation when a level raises
    above the surface of the Dirac sea.\label{LevelCrossing}}
\end{figure}

At the perturbative level, there are four conserved fermion numbers in
the SM: The baryon number $B$ and three lepton numbers $L_\alpha$. The
observed neutrino oscillations, which cannot be explained within the
SM, clearly violate the individual lepton numbers $L_\alpha$. If
neutrinos are Majorana particles (as e.g. in the model presented in 
section \ref{nuMSM}), the Majorana mass term also violates the total
lepton number $L=\sum_\alpha L_\alpha$.  Baryon number $B$ is
conserved at each order in perturbation theory, but violated by
non-perturbative effects \cite{'tHooft:1976up}. This can be seen by
looking at the baryonic current $j_\mu^B$.  $j_\mu^B$ is conserved
classically, but obtains a non-zero divergence by a quantum anomaly
\cite{Adler:1969gk},
\begin{equation}
\partial^\mu j_\mu^B=\frac{n_f}{32\pi^2}\big(-g^2{\rm tr}
({\rm F}_{\mu\nu}\tilde{{\rm F}}^{\mu\nu})
+g'^2{\rm F'}_{\mu\nu}\tilde{{\rm F'}}^{\mu\nu}\big),
\end{equation}
where $g$, ${\rm F}_{\mu\nu}$ and $g'$, ${\rm F'}_{\mu\nu}$ are the
gauge coupling and field strength tensor of the SU(2) and U(1) gauge
interaction, respectively, and $n_f=3$ is the number of fermion
families. Baryon number violation is most easily explained
semi-classically by {\it fermionic level crossing}
\cite{Callan:1976je,Jackiw:1976pf}. In the bosonic sector of the
standard electroweak theory, there is an infinite number of field
configurations that minimise the static energy functional, which we
refer to as ``vacua''. They are physically equivalent, but can be
distinguished by the Chern-Simons number $N_{CS}$ of the gauge field
configuration.   The energy levels of fermions depend on the bosonic
background fields. Fermionic level crossing occurs when $N_{CS}$
changes; then an energy level raises above (or falls below) the
surface of the Dirac sea, see figure \ref{LevelCrossing}, which means
that fermions are created (or absorbed) by the background. All SU(2)
doublets are subject of level crossing, leading to the simultaneous
creation (or disappearance) of $9$ quarks and $3$ leptons. Thus,
$L_\alpha-\frac{B}{3}$ remain conserved in sphaleron processes though
$B$ and $L_\alpha$ are violated individually. The different vacua are
separated by a potential barrier, the height of which can be estimated
by the {\it sphaleron} energy $M_{\rm sph}\sim M_W/\alpha_W$. The
sphaleron is the field configuration of maximal energy along the path
of minimal energy in field space that connects two minima in the
classical potential \cite{Klinkhamer:1984di}.

In quantum field theory, the early universe can be described as a
thermodynamic ensemble characterised by a density matrix
$\varrho=\exp(-{\rm H}/T)$, where ${\rm H}$ is the Hamiltonian. At low
temperatures $T\ll T_{EW}$, only configurations with field expectation
values near the minima of the effective potential are significantly
populated. Tunneling through the potential barrier is the only
process that can change $N_{CS}$ from such an initial state. The
tunneling rate at $T=0$ \cite{'tHooft:1976up} is suppressed by
$\exp(-\frac{4\pi}{\alpha_W})\sim 10^{-160}$, effectively forbidding
$B$-violation in vacuum. Near $T\sim T_{EW}$ the Boltzmann suppression
$\exp(-{\rm H}/T)$ is less efficient and $N_{CS}$ can change by a
classical transition due to thermal fluctuations.  The probability for
these baryon number violating processes is governed by the sphaleron
rate \cite{Khlebnikov:1988sr}
\begin{equation}
\Gamma_{\rm sph}\equiv\lim_{t,V\rightarrow\infty}
\frac{(N_{CS}(t)-N_{CS}(0))^2}{V t}=\int d^4x 
\langle \partial^\mu j_\mu^B(x) \partial^\nu j_\nu^B(0) \rangle,
\end{equation}
where $V$ is the total volume. The (finite temperature) effective
potential and sphaleron configuration change in the vicinity of the
electroweak symmetry restoration.  In the Higgs phase, $\Gamma_{\rm
sph}$ is given by
\begin{equation}\label{broken}
\Gamma_{\rm sph}=A\ (\alpha_W T)^4 \Big(\frac{M_{\rm
sph}}{T}\Big)^7\exp\Big(-\frac{M_{\rm sph}}{T}\Big)\ \ \ {\rm( \ Higgs
\ phase)},
\end{equation}
where $A$ is a coefficient that can be determined numerically
\cite{Arnold:1987mh,Kunz:1992uh,Moore:1995jv,Akiba:1989xu,
Carson:1989rf,Carson:1990jm,Baacke:1993aj,Arnold:1992rz,
Fodor:1994bs,Farakos:1994kx,Arnold:1996dy,Moore:1998swa}.
For $T\gg T_{EW}$, in the symmetric phase, the Boltzmann suppression
is cancelled because $M_W$ in $M_{\rm sph}$ is replaced by the
non-abelian magnetic screening scale  $\sim \alpha^2 T$ and
$\Gamma_{\rm sph}$ reads \cite{Bodeker:1999gx} (see also
\cite{Shanahan:1998gj})
\begin{equation}\label{symm}
\Gamma_{\rm sph}=(25.4\pm 2.0)\alpha_W^5 T^4 \ \ \ 
{\rm (\ symmetric \ phase\ )},
\end{equation}
with $\alpha_W=g^2/4\pi$. A more refined expression for general
$SU(N)$ can be found in \cite{Moore:2010jd}. 
Recently the sphaleron rate has been calculated numerically throughout the electroweak transition, interpolating between (\ref{broken}) and (\ref{symm})  \cite{D'Onofrio:2012jk}.
For Higgs masses between
$100$ GeV and $300$ GeV, sphaleron reactions in the SM become slower
than the rate of the universe's expansion at a temperature $T_{\rm
sph}$, with $100$ GeV $\lesssim T_{\rm sph}\lesssim 300$ GeV
\cite{Burnier:2005hp}. Above this temperature, baryon number is
efficiently violated and all Sakharov conditions are fulfilled. Thus,
the SM in principle can provide a framework for baryogenesis. 

However, it turns out that both, the violation of $CP$, condition II),
and the deviation from thermal equilibrium, condition III), are not
large enough to produce the observed $\eta$ (and thus $\Omega_B$).  
The amount of CP-violation can be estimated by constructing
reparametrisation invariant objects out of the quark mass matrices
\cite{Shaposhnikov:1987pf,Brauner:2011vb}. The lowest order
CP-noninvariant combination is the Jarlskog determinant
\cite{Jarlskog:1985ht}, which in terms of quark masses and mixing
angles reads
\begin{eqnarray}
D&=&\sin(\theta_{12})\sin(\theta_{23})\sin(\theta_{13})\delta_{KM}\\
&&(m_t^2-m_c^2)(m_t^2-m_u^2)(m_c^2-m_u^2)(m_b^2-m_s^2)
(m_b^2-m_d^2)(m_s^2-m_d^2)\nonumber
.\end{eqnarray}
A dimensionless quantity can be constructed when dividing by the
relevant temperature $T_{\rm sph}$, at which the BAU freezes out, to
the $12^{th}$ power, $D/T_{\rm sph}^{12}\sim 10^{-20}\ll \eta$. Though
not a direct proof of impossibility, the smallness of this result
makes baryogenesis within the SM very challenging
\cite{Shaposhnikov:1987pf,Shaposhnikov:1987tw,Ambjorn:1988gf,Farrar:1993hn,
Brauner:2011vb}.

During most of the history of the universe, cosmic expansion is the
only source of non-equilibrium. Within the SM, the BAU has to be
created around $T\sim T_{\rm sph}$,  as otherwise it would be washed
out by sphaleron processes. At these temperatures, all particle
reactions in the SM act much faster than cosmic expansion (their rates
are much larger than the Hubble parameter), keeping all particle
species very close to thermal equilibrium.   The only way to cause a
significant deviation from equilibrium at $T\sim T_{\rm sph}$,
necessary to satisfy condition III) \cite{Dolgov:1979mz}, would be a
first order phase transition from the symmetric to the Higgs phase of
the electroweak theory \cite{Shaposhnikov:1987tw}. It proceeds  via
nucleation of bubbles of new (Higgs) phase. The  bubbles expand
rapidly, as the field configuration inside is energetically more
favourable. The mechanism of baryogenesis related to bubble wall
expansion  is based on the following picture \cite{Nelson:1991ab} (for
spinodial decomposition phase transition see \cite{McLerran:1990zh}).
When the bubble wall, which separates the symmetric phase from the
Higgs phase, passes through the medium, it can cause a large
deviation from equilibrium. Due to the CP-violation, the reflection
and transmission coefficients for quarks and antiquarks colliding with
the bubble wall are different; this allows to generate a
matter-antimatter asymmetry, which can dissipate into the bubble. It
is preserved in the Higgs phase from washout because sphalerons are
inefficient, but disappears in the symmetric phase, where baryon
number non-conservation is rapid.    

In the electroweak theory the symmetric and the Higgs phases are
continuously connected. A first order phase transition could only
occur if the Higgs mass were below $72$ GeV
\cite{Kajantie:1996mn,Rummukainen:1998as,Csikor:1998eu}; this is in
clear contradiction with experimental data
\cite{ATLAS:2012ae},  allowing to conclude that it is
extremely unlikely that the observed BAU can be generated within the
SM.

In the past decade, two clear signs of particle physics beyond the SM
other than the BAU have been found experimentally. These are the
discovery of neutrino flavour changing processes, usually interpreted
as oscillations\footnote{For a comprehensive review and references to original experimental and theoretical work see \cite{Strumia:2006db}}, and the conclusion that the
observed DM cannot be baryonic. The latter is based on BBN and CMB
precision data \cite{Fields:2006ga,Komatsu:2010fb}, which in
combination with accurate measurements of the Hubble parameter
\cite{Riess:2009pu} and simulations of structure formation show that
the amount of matter in the universe exceeds the amount of baryonic
matter by a factor $\sim 6$ \cite{Lahav:2010mi}. The explanation of
these experimental facts unavoidably requires physics beyond the SM.
Extensions of the SM generally contain new sources of CP violation and
often also $B$ violation, and there is not much motivation to insist
on a source for the BAU within the SM. In section \ref{nuMSM} we
explore the possibility that neutrino oscillations and the observed DM
have a common origin that is also responsible for the BAU.

\subsection{Beyond the SM}
Baryogenesis necessarily involves $B$-violating processes due to
condition I). At the same time, the current non-observation of these,
e.g. in proton decay or $n-\bar{n}$-oscillations, strongly constrains
$B$-violation in the present day universe. An enormous number of
models that are in accord with these conditions have been suggested
since Sakharov. Often they are grouped into those that directly
generate a matter-antimatter asymmetry in the baryonic sector and
those that initially generate a lepton asymmetry (``leptogenesis''),
which is then transferred to the baryonic sector, either by SM
sphalerons or processes that involve physics beyond the SM. 

One can alternatively classify the variety of models into {\it
top-down} and {\it bottom-up} approaches.  In top down approaches, the
underlying theory has usually not been developed for the purpose to
explain the BAU,  but is motivated by more general theoretical or
aesthetic considerations. The most prominent examples are grand
unified theories (GUT) and supersymmetry (SUSY), but also string
inspired scenarios.  The requirement to predict the correct BAU is a
necessary condition that can be used to constrain the parameter space
for these classes of models. 

Bottom-up approaches, on the other hand, take the SM as a basis and
add ingredients that allow to explain the BAU.  Ideally, these account
also for other phenomena that cannot be explained within the SM. A
guideline for the exploration of the infinite-dimensional space of
possible SM-extensions can be the principle of minimality ("Ockhams
razor"), often accompanied by ``naturalness'' considerations. We
discuss a specific model that obeys these principles in section
\ref{nuMSM}.   Bottom-up models may be viewed as effective field
theories, without knowledge of the underlying physics at higher energy
scales. 

Though theoretically very interesting, most models of baryogenesis are
hard to falsify.  They may remain viable even if no signals are
observed in any experiments in the centuries to come, as they can be
saved from falsification by pushing up an associated energy scale. In
some cases, indirect evidence that supports the underlying theory may
be found in low energy experiments or astrophysical observations, but
even then it would be unlikely that these data single out one model
and the source of the BAU can be identified uniquely. In foreseeable
time, only three of the popular scenarios are testable in the strict
sense that all parameters of the theory may be measured: electroweak
baryogenesis,  resonant leptogenesis \cite{Pilaftsis:1997jf} and
baryogenesis from sterile neutrino oscillations
\cite{Akhmedov:1998qx,Asaka:2005pn}. The former two are covered in
other parts \cite{Morrissey:2012db,lepto} of \cite{FocusIssue}, we therefore
in the following only discuss baryogenesis from neutrino oscillations.

\subsection{Baryogenesis from Sterile Neutrino Oscillations}
\label{nuMSM}
In the SM, neutrinos are the only fermions that appear only as left
chiral fields.  At the same time, neutrino flavour changing processes,
usually interpreted as oscillations, cannot be explained within the
model. Complementing the SM by right handed neutrinos that are singlet under
all gauge interactions, offers an attractive explanation for neutrino
oscillations due to masses generated by the seesaw mechanism
\cite{Minkowski:1977sc}.  This model is described by the Lagrangian
\begin{eqnarray}
	\label{nuMSM_lagrangian}
	\mathcal{L}_{\nu MSM} &=&\mathcal{L}_{SM}+ i
\bar{\nu_{R}}\slashed{\partial}\nu_{R}-\bar{L_{L}}F\nu_{R}\tilde{\Phi}
-\bar{\nu_{R}}F^{\dagger}L_L\tilde{\Phi}^{\dagger} \nonumber\\ 
&&-{\rm \frac{1}{2}}(\bar{\nu_R^c}M_{M}\nu_{R}
+\bar{\nu_{R}}M_{M}^{\dagger}\nu_{R}^c),
	\end{eqnarray}
where we have suppressed flavour and isospin indices.
$\mathcal{L}_{SM}$ is the Lagrangian of the SM. $F$ is a matrix of
Yukawa couplings and $M_{M}$ a Majorana mass term for the right handed
neutrinos $\nu_{R}$. $L_{L}=(\nu_{L},e_{L})^{T}$ 
are the left handed lepton doublets and $\Phi$ is the Higgs doublet.

The Lagrangian (\ref{nuMSM_lagrangian}), with eigenvalues of $M_M$ far
above the electroweak scale, is the basis of thermal leptogenesis
scenarios \cite{Fukugita:1986hr}, in which the CP-asymmetry
responsible for the BAU is generated during the freezeout and decay of
right handed neutrinos. The attractive feature of this setup is that
it provides a common explanation for the small neutrino masses and the
BAU within GUTs.

For eigenvalues of $M_M$ below the electroweak scale, the Lagrangian
(\ref{nuMSM_lagrangian})  yields the possibility that the asymmetry
was created during the thermal production of right handed (sterile)
neutrinos in the early universe - rather than during their freezeout
and decays \cite{Akhmedov:1998qx,Asaka:2005pn}.  This mechanism was
called baryogenesis via (sterile) neutrino oscillations in
\cite{Akhmedov:1998qx}. It is, however, also efficient when the
oscillations are practically not relevant because they are e.g. too
rapid and average out. The crucial point is that the initial sterile
neutrino abundance deviates from its equilibrium value, and chemical
equilibrium is not established before sphaleron freezeout. 

This possibility is realised in the {\it Neutrino minimal Standard
Model} ($\nu$MSM), which can be viewed as a minimal extension of the
SM. This in particular means that there is no modification of the
gauge group, the number of fermion families remains unchanged and no
new energy scale above the Fermi scale is introduced. It can explain
simultaneously three empirical facts that cannot be understood within
the framework of the SM: The observed neutrino oscillations, dark
matter and the BAU. Here we focus on the latter. An introduction to
the $\nu$MSM, along with the most recent bounds on its parameter
space, can be found in \cite{Canetti:2012kh}, for additional reading see
\cite{Boyarsky:2009ix,Shaposhnikov:2008pf,Asaka:2005pn,
Canetti:2010aw,Asaka:2010kk,Asaka:2011wq,letter}. 

The Lagrangian (\ref{nuMSM_lagrangian}) yields six different neutrino
mass eigenstates. Three of them are mixes of the ``active'' SM
neutrinos $\nu_\alpha$ ($\alpha=e,\mu,\tau$) with masses $m_i$. The
other three are ``sterile'' neutrinos $N_1$, $N_2$ and $N_3$ with
masses $M_I$.   Mixing between active and sterile neutrinos is
suppressed by small angles $\theta_{\alpha I}=(m_D M_M^{-1})_{\alpha
I}$, where $m_D=Fv$ and $v$ is the Higgs expectation value. The mass
matrix $m_\nu$ for the active neutrinos, leading to the observed
neutrino oscillations, is generated by the seesaw mechanism
\cite{Minkowski:1977sc,Fukugita:1986hr}; $m_\nu\simeq-\theta
M_M\theta^T$. For eigenvalues of $M_M$ below the electroweak scale,
this requires the Yukava couplings $F$ to be tiny.

The requirement to produce the correct BAU can be used to constrain
the $\nu$MSM parameter space and find the experimentally interesting
region.  In the following we assume that only two sterile neutrinos
$N_{2,3}$ participate in baryogenesis, which leaves open the
possibility to use $N_1$ as DM candidate. In the simplest scenario,
$N_{2,3}$ are not produced during reheating due to their tiny Yukawa
interactions \cite{Bezrukov:2008ut}. Instead, they are produced
thermally from the primordial plasma during the radiation dominated
epoch.  Since they are generated as flavour eigenstates, they undergo
oscillations.  Throughout this nonequilibrium process, all Sakharov
conditions are fulfilled and baryogenesis is possible if several
requirements are fulfilled. On one hand, the (temperature dependent)
mass splitting $|\delta M|=|M_3-M_2|/2$ has to be large enough for
the  neutrinos to perform several oscillations, on the other hand it
has to be small to ensure resonant amplification. Finally, $N_{2,3}$
should not reach chemical equilibrium before sphaleron freezeout to
avoid washout. 

If the $\nu$MSM shall, apart from the BAU, also explain the observed
DM density $\Omega_{DM}$, the lightest sterile neutrino ($N_1$) is
required to be sufficiently long lived to constitute all DM. Then its
mixing angle must be too small to be seen in collider experiments. However,
being a decaying DM candidate, it can be searched for indirectly by
astronomical observations \cite{Boyarsky:2009ix}.  These, together
with the seesaw formula, constrain the $N_1$ mass to $1$ keV $\lesssim
M_1 \lesssim 50$ keV.  This implies that one active neutrino is
effectively massless, which fixes the absolute scale of neutrino
masses. In contrast, $N_{2,3}$ can be seen directly at collider
experiments \cite{Gorbunov:2007ak}. Another consequence of the small
$N_1$ mixing is that its coupling is too small to contribute to the
generation of a BAU.  Baryogenesis can therefore be described in an
effective theory with only two sterile neutrinos $N_{2,3}$.
CP-violating oscillations amongst them can generate a lepton
asymmetry, which can be translated into a BAU by SM sphalerons.  It
turns out that for  $N_{2,3}$ masses $M_{2,3}=\bar{M}\mp\delta M$ in
the GeV range, which are accessible to direct search experiments, the
observed BAU can only be generated when the masses are
quasi-degenerate, i.e. $\delta M\ll \bar{M}$. The degeneracy is
essential for a resonant amplification of the CP-violating effects.  
If one, however, drops the requirement that the Lagrangian
(\ref{nuMSM_lagrangian}) shall also explain the observed DM, all three
sterile neutrinos can participate in baryogenesis. It has been found
in \cite{Drewes:2012ma} that in this case, no mass degeneracy is
needed due to the additional sources of CP-violation in the couplings
of $N_1$.

The parameter space can be studied in a quantitative manner by means
of effective kinetic equations, similar to those commonly used in
neutrino physics \cite{Sigl:1992fn,Asaka:2011wq}. They allow to track
the time evolution of the $N_I$ and lepton chemical potentials.
Detailed studies have been performed in
\cite{Canetti:2010aw,letter,Canetti:2012kh}.  During this calculation one
can approximate $M_1=0$ and drop $N_1$ from the Lagrangian, which has
no effect on baryogenesis. This leaves $11$ parameters in addition to
the SM. In the Casas-Ibarra parametrisation for $F$
\cite{Casas:2001sr}, these are three active mixing angles, two active
neutrino masses, two Majorana masses $M\pm\Delta M$ in $M_M$, one
Dirac phase, one Majorana phase and the real and imaginary part of a
complex angle $\upomega$.  

\begin{figure}[!h]
\centering
\begin{tabular}{cccc}
\includegraphics[width=7.5cm]{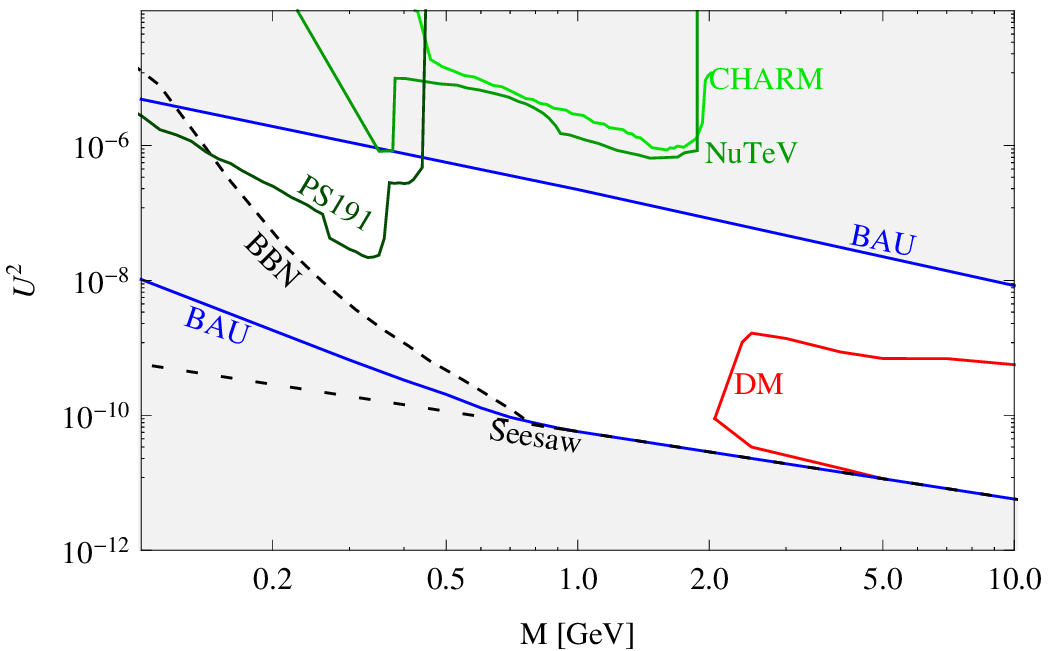} &
\includegraphics[width=7.5cm]{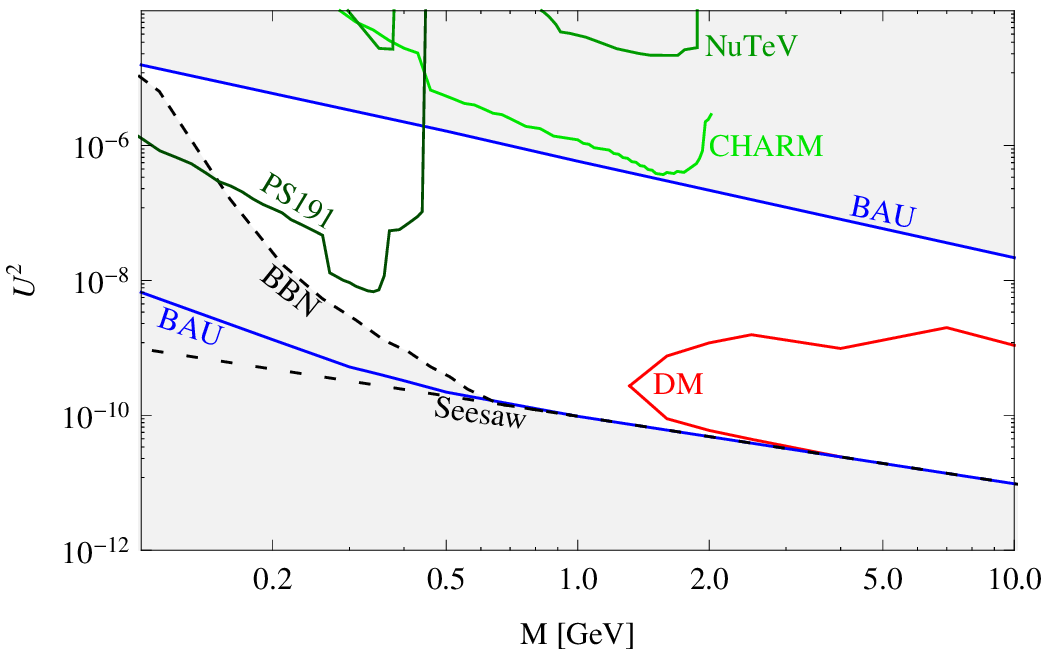} 
\end{tabular}
\caption{Constraints on sterile neutrino mass $M$ and mixing $U^2={\rm
tr}(\theta^\dagger \theta)$ in the $\nu$MSM as found in
\cite{Canetti:2012kh,letter} for normal (left panel) and inverted (right
panel) hierarchy of active neutrino masses. In the regions below the
black dashed "seesaw" line there exists no choice of $\nu$MSM
parameters that is in accord with experimental constraints on the
active neutrino mixing matrix. In the region below the black dotted
BBN line, the lifetime of $N_{2,3}$ particles in the early universe is
larger than $0.1$s, yielding the danger that their decay spoils the
agreement between BBN calculations and observed light element
abundances. The regions above the green lines of different shade are
excluded by the NuTeV \cite{Vaitaitis:1999wq}, CHARM
\cite{Bergsma:1985is} and CERN PS191 \cite{Bernardi:1985ny}
experiments, as indicated in the plot. In the region between the blue
lines, a CP-asymmetry that explains the observed BAU can be produced
during the thermal production of $N_{2,3}$. In the region within the
red line, thermal production of $N_1$ (resonant and non-resonant) is
sufficient to explain all observed DM. The CP-violating phases that
maximise the efficiency of baryogenesis and DM production are
different. They were chosen independently for the blue and red line
displayed here.  The region in which $\Omega_B$  and $\Omega_{DM}$ can
be explained {\it simultaneously} almost coincides with the area
inside the red line, see \cite{Canetti:2012kh}.\label{MU2}}
\end{figure}

Figure \ref{MU2} shows the sterile neutrino masses $\bar{M}\simeq M$
and mixings $U^2={\rm tr}(\theta^\dagger\theta)$ for which the
observed BAU can be generated by the two sterile neutrinos $N_{2,3}$.
It also displays other experimental bounds and constraints from BBN. 
To obtain these results, all known parameters have been fixed to their
experimental values found in \cite{Fogli:2011qn}, while the
CP-violating phases have been chosen to maximise the lepton
asymmetries. 

If one requires the $N_1$ density to make up for the observed
$\Omega_{DM}$, its thermal production rate needs to be amplified
resonantly \cite{Shi:1998km}. The resonant amplification is due to a
level crossing between active and sterile neutrino dispersion
relations, caused by the MSW effect
\cite{Wolfenstein:1977ue,Mikheev:1986gs},  and requires the presence
of a lepton asymmetry $|N_L-N_{\bar{L}}|/s\gtrsim 8\cdot 10^{-6}$ in
the plasma \cite{Laine:2008pg}. This LAU, which exceeds the BAU by
orders of magnitude, has to be produced during the freezeout and decay
of $N_{2,3}$. Such large asymmetry can only be generated from
$N_{2,3}$ oscillations if their masses are highly degenerate, with a
physical mass splitting $\delta M$  of the order of active neutrino
masses $m_i$ \cite{letter}. This degeneracy is much stronger than the
one required to explain the BAU. Thus, the BAU and DM production in
the $\nu$MSM are both related to lepton asymmetries in the early
universe, produced by the same sources of CP-violation. This may hint
to understand the similarity $\Omega_{DM}\sim\Omega_B$, though it does
not provide an obvious explanation because today's values of
$\Omega_B$ and $\Omega_{DM}$ depend on several other parameters.  The
requirement to generate a sufficient LAU allows to impose further
constraints on the sterile neutrino properties. This was included in
the analysis in \cite{letter,Canetti:2012kh}. The computational effort for
quantitative studies is huge, as it requires to track time evolution
from hot big bang initial conditions down to temperatures $\sim 100$
MeV, below hadronisation. At different temperatures, different
processes enter the effective Hamiltonian. Various time scales,
related to production, oscillations, decoherence, freezeout and decay
of the $N_I$ are involved. A detailed study has been performed in
\cite{Canetti:2012kh,letter}.

Currently, the main uncertainty in these results comes from the
kinetic equations. The BAU is generated from a quantum interference,
in a regime where coherent oscillations can be essential.   The
semiclassical kinetic equations known from neutrino physics most
likely capture the main features of this process, but may require
corrections in the resonant regime. A first principles derivation
\cite{Garny:2011hg,Garbrecht:2011aw} is required to determine the size
of these. This is not entirely specific to the $\nu$MSM; a consistent
description of transport phenomena involving quantum interference,
flavour effects and CP-violation remains an active field of research
in many scenarios of baryogenesis
\cite{Buchmuller:2000nd,Cline:2000nw,Prokopec:2003pj,De
Simone:2007rw,Anisimov:2008dz,Garny:2009qn,Cirigliano:2009yt,Beneke:2010dz,Herranen:2010mh,
Anisimov:2010dk,Garbrecht:2011aw,Gagnon:2010kt,Garny:2011hg,Garbrecht:2011xw,
Garbrecht:2012qv}. 
\section{Conclusions}
The origin of matter remains one of the great mysteries in physics.
Observationally we can be almost certain that the present day universe
contains no significant amounts of (baryonic) antimatter, and the
baryons are the remnant of a small matter-antimatter asymmetry $\sim
10^{-10}$ in the early universe. This asymmetry cannot be explained
within the Standard Model of particle physics and cosmology. It
provides, along with neutrino oscillations, dark matter and
accelerated cosmic expansion, one of the few observational proofs of
physics beyond the SM.  Many extensions of the SM are able to explain
the BAU. However, since it is characterised by only one observable
number, the BAU cannot be used to pin down the correct model realised
in nature. In spite of that, it provides a necessary condition that
can be used to exclude or constrain models, and baryogenesis remains a
very active field of research. Amongst the many suggested theories of
baryogenesis, only a few are experimentally testable.  We discussed an
example of this kind, in which right handed neutrinos are the common
origin of the BAU, DM and neutrino oscillations.  
\\ 
\newline
{\large \textbf{Acknowledgements}}
This work was supported by the Swiss National Science Foundation and
the Gottfried Wilhelm Leibniz program of the Deutsche
Forschungsgemeinschaft. MD would like to thank Signe Reimer-S\o
rensen, Yvonne Wong, Maik Stuke and John Webb for helpful comments.

\begin{small}

\end{small}
\end{document}